# Preliminary Design of The Wide-Field Infrared Survey Explorer (WISE)

Amanda K. Mainzer[1], Peter Eisenhardt[1], Edward L. Wright[2], Feng-Chuan Liu[1], William Irace[1], Ingolf Heinrichsen[1], Roc Cutri[3], Valerie Duval[1],
[1]Jet Propulsion Laboratory, 4800 Oak Grove Dr., Pasadena, CA USA 91109
[2]University of California, Los Angeles
[3]Infrared Processing and Analysis Center, 770 South Wilson Ave., Pasadena, CA USA 91125

**ABSTRACT**

The Wide-field Infrared Survey Explorer (WISE), a NASA MIDEX mission, will survey the entire sky in four bands from 3.3 to 23 microns with a sensitivity 1000 times greater than the IRAS survey. The WISE survey will extend the Two Micron All Sky Survey into the thermal infrared and will provide an important catalog for the James Webb Space Telescope. Using $1024^2$ HgCdTe and Si:As arrays at 3.3, 4.7, 12 and 23 microns, WISE will find the most luminous galaxies in the universe, the closest stars to the Sun, and it will detect most of the main belt asteroids larger than 3 km. The single WISE instrument consists of a 40 cm diamond-turned aluminum afocal telescope, a two-stage solid hydrogen cryostat, a scan mirror mechanism, and reimaging optics giving 5" resolution (full-width-half-maximum). The use of dichroics and beamsplitters allows four color images of a 47'x47' field of view to be taken every 8.8 seconds, synchronized with the orbital motion to provide total sky coverage with overlap between revolutions. WISE will be placed into a Sun-synchronous polar orbit on a Delta 7320-10 launch vehicle. The WISE survey approach is simple and efficient. The three-axis-stabilized spacecraft rotates at a constant rate while the scan mirror freezes the telescope line of sight during each exposure. WISE is currently in its Preliminary Design Phase, with the mission Preliminary Design Review scheduled for July, 2005. WISE is scheduled to launch in mid 2009; the project web site can be found at www.wise.ssl.berkeley.edu.

## 1. INTRODUCTION

The Wide-Field Infrared Survey Explorer (WISE) will perform an all-sky survey a 4 infrared wavelengths: 3.3, 4.7, 12 and 23μm. It will utilize 1024x1024 detector arrays which reach sensitivity limits of 120, 160, 650, and 2600 μJy at 3.3, 4.7, 12, and 23μm. (These bands are denoted 1, 2, 3, and 4.) The facility carries a 40 cm telescope providing a resolution of 5" (10" at 23μm) with a 47 arcmin instantaneous field of view. The operating temperatures will be 30–34K for the 3.3 & 4.7 μm detectors, 7.8 ± 0.5K for the 12 & 23μm detectors and 17K for the optical system, which are achieved using a two stage solid hydrogen cryostat providing a minimum mission lifetime of 7 months allowing for a single full coverage of the entire sky. Figure 1 depicts the WISE spacecraft. The instrument takes images continuously every 11 sec, while the spacecraft maintains a continuous pitch rate that matches the orbit pitch. As the Observatory is scanning the sky, the sun-synchronous 500 km altitude orbit

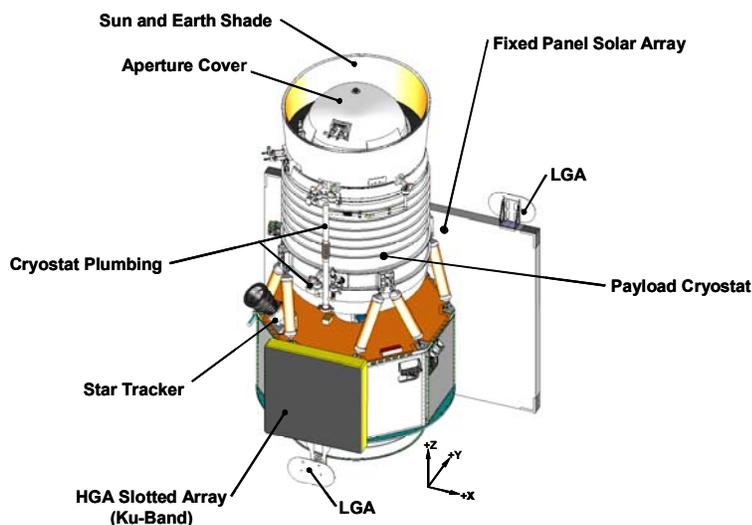

**Figure 1: The WISE Observatory.**

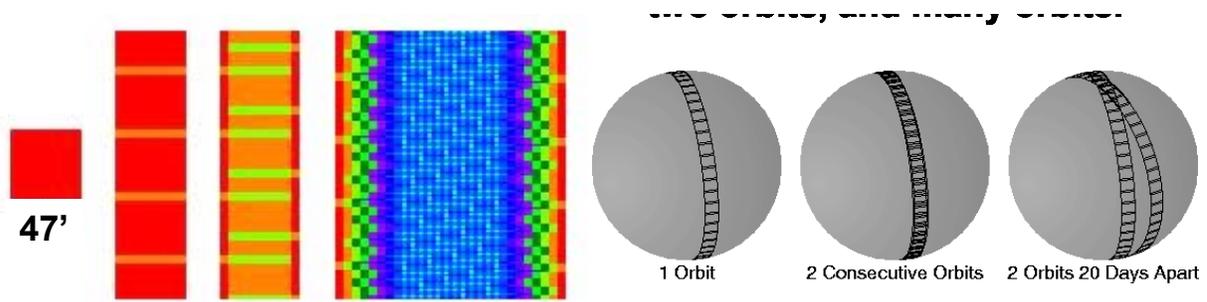

**A single frame, a single orbit, two orbits, and many orbits.**

**Figure 2:** The WISE survey strategy, showing how the entire sky can be surveyed in 6 months with a 47' field of view.

processes around the celestial sphere in 6 months. A scan mechanism offsets the orbital motion and freeze the sky on the arrays. Figure 2 depicts the progress of the WISE survey. After allowing for losses due to the moon and South Atlantic Anomaly (SAA), the minimum number of exposures per position is 4 with a median of 14.

The final output of the WISE mission will be an all-sky catalog. The raw data from the mission of about 50 Gbytes/day (24 Gbytes/day compressed using a lossless Rice algorithm which provides 2.1:1 compression) will be processed to provide a source catalog and image atlas. The WISE Observatory will be launched by a Delta 7320-10 rocket in June, 2009. Table 1 summarizes the WISE implementation. WISE will be co-manifested with the NASA New Millennium mission Space Technology 8 (ST8); ST8's orbit and launch requirements are compatible with WISE.

| Parameter | Value |
|---|---|
| Orbit | 500km; 97.3° inclination; sun synchronous 6am/6pm |
| Orbit Period | 94.6 min. (500 km circular orbit) |
| # of Frames/Orbit | 516 frames/orbit at 11 sec cadence |
| Frame-to-Frame Overlap | 10% |
| Orbit-to-Orbit Overlap | 90% |
| Launch Vehicle | Delta 7320-10; 750 kg to 500 km polar orbit; 10' fairing |
| Mass-Total | 750 kg (32% margin) |
| Power-Total | 467 W EOL capability |
| Number of Repeats | 4 minimum; 14 median |
| Deployed Solar Panels | No; 3.0 m$^2$ |
| Fixed HGA | Yes – slotted |
| Telecom Service | TDRSS Ku-band for science downlink; TDRSS S-band for command & real time engineering data downlink |
| Number of Contacts/Day | 4 (up to 60 min. total) |
| Telescope Diameter | 40 cm |
| F/number | 3.375 |
| Angular Pixel Size | 2.75 arcsec IFOV; 46.9 arcmin FOV |
| Scan Mirror | Yes |
| Number of Detectors | 4; 2 HgCdTe, 2 Si:As |
| Detector Format | 1024 x 1024; 18 μm pixels |
| Binning | Yes, 23 μm band has 2 x 2 binning |
| Integration Time | 8.8 sec out of 11 sec cadence (9 samples up the ramp) |
| Data Rate | TBR Mbps to S/C including binning<br>51 GB/day uncompressed with headers<br>24 GB/day compressed (2.1:1 lossless) |
| Cryostat Type | 2-stage solid hydrogen |
| Cryostat Lifetime | 13 months + 3.25 months margin |

**Table 1: The WISE implementation.**

## 2. WISE Science

We live in an era when the basic reconnaissance of the universe is underway. Sensitive all-sky surveys across the electromagnetic spectrum are imminent or have recently been completed, but the mid-infrared (IR) lags behind. Figure 2a shows a moon-sized patch of the sky seen with the mid-IR sensitivity and angular resolution of the Infrared Astronomical Satellite (IRAS), the best existing all-sky survey at these wavelengths. Vast advances in IR detector technology have occurred since the IRAS survey was completed in 1983, enabling the WISE survey. WISE will have 1,000 times better sensitivity than IRAS at 12 and 23 μm (Figure 2b), and 500,000 times better sensitivity than the Cosmic Background Explorer (COBE) at 3.3 and 4.7 μm. Every survey with such a leap in sensitivity and covering much or all of the sky has revealed dramatic surprises, which have significantly changed our knowledge of our universe and have created entire new areas of astronomical investigation. Examples include galaxy clusters from the Palomar Observatory Sky Survey (POSS), quasars from the 3C radio survey, planetary debris disks from IRAS, and L and T-dwarf stars from 2MASS. WISE will deliver to the scientific community over one million calibrated rectified images covering the whole sky, and catalogs of half a billion objects, in these four bands. The wavelength range of WISE largely overlaps that planned for the James Webb Space Telescope (JWST). The power of all-sky surveys can be illustrated by examining the citation rate of IRAS data, which has remained unchanged since its launch in 1983 (see Figure 3). WISE's primary goal is to fill the wavelength gap between the 2MASS all-sky survey at 1.2 – 2.2 μm and the Astro-F all-sky survey at 50 and 100 μm (see Figure 4).

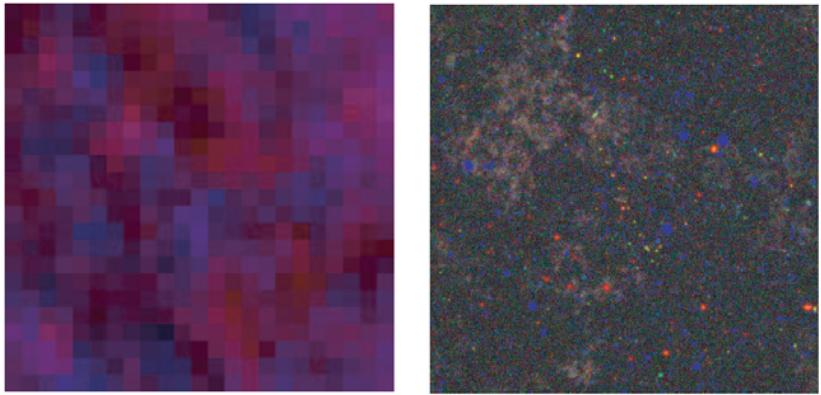

**Figure 2a: 34 arcmin wide region in the mid-IR seen with IRAS's sensitivity and resolution. Figure 2b: The same region seen by WISE.**

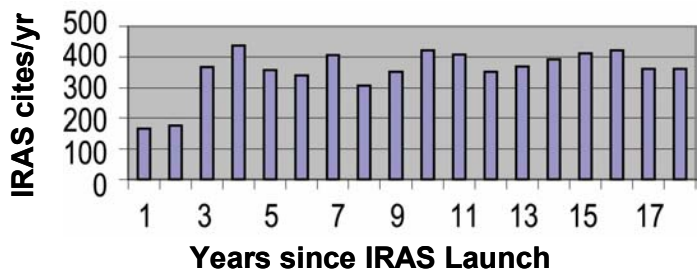

**Figure 3: The value of all-sky surveys: IRAS citations per year remain unchanged since its launch >20 years ago.**

With a sensitive all-sky survey in four bands, WISE is optimized to achieve its two primary science goals, which are to

- Study the nature and evolutionary history of ultra-luminous IR galaxies (ULIRGs), and identify the most luminous galaxies in the universe; map clusters of galaxies and large scale structure within 7-billion light-years; and

- Measure the space density, mass function, and formation history of brown

dwarf (BD) stars in the solar neighborhood identifying the closest stars to the sun.

In addition, WISE will address other fundamental topics in astrophysics, including

- Determining the radiometric albedos for almost all known asteroids (>>100,000);

- Measuring the very faint end of the luminosity function of protostars in nearby star formation regions

- Contributing to the understanding of the evolution of circumstellar disks

Table 2 summarizes the basic scientific requirements for the WISE mission. These include the sensitivity, photometric accuracy, catalog reliability, sky coverage, number of repeat observations at each point in the sky, astrometric accuracy, and catalog completeness. The sensitivity requirement was set by the need to detect WISE's two primary science goals: finding the nearest star to our Sun and finding the most luminous galaxies in the universe. The photometric accuracy requirement was driven by the need to accurately distinguish different object types from each other by the use of color-color diagrams. The minimum number of repeat observations was driven by the need to protect against spurious sources such as cosmic rays. Catalog reliability and completeness ensure that sources in the catalog are real, and that all real sources are actually detected, respectively. Photometric accuracy, completeness, and reliability down to SNR = 5 will be characterized for the entire survey.

| Requirement | Implementation |
| --- | --- |
| Image atlas | Digital image atlas combining multiple exposures at each sky position |
| Source catalog | Catalog of sources associated with image atlas. Include sources to SNR 5 in any band, characterize completeness and reliability at all flux levels |
| Minimum # of bands | Four bands centered within 10% of 3.3, 4.7, 12,and 23 microns |
| Band 1 bandpass | 2.8 – 3.8 μm |
| Band 2 bandpass | 4.1 – 5.2 μm |
| Band 3 bandpass | 7.5 – 16.5 μm |
| Band 4 bandpass | 20 – 28 μm* |
| Number of repeat observations at each point in sky | At least 4 independent exposures in each filter over at least 95% of the sky |
| SNR = 5 Sensitivity | 0.12/0.16/0.65/2.6 mJy in bands 1, 2, 3, 4 assuming 8 repeats |
| Reliability | Reliability > 99.9% for SNR > 20 |
| Completeness | Completeness > 95% for SNR > 20 |
| Photometric accuracy | <7% relative photometric accuracy for SNR > 100 |
| Astrometric accuracy | Position error <0.5" with respect to 2MASS for SNR > 20 |
| Saturation flux density |  |
| Public data release | Image atlas and catalog publicly available within 17 months of end of on-orbit data collection |

**Table 2. WISE science requirements. *Band 4 cutoff is determined by the detector long wavelength cutoff.**

## 2.1. Brown Dwarfs

A major goal for WISE is to identify brown dwarfs closer than the nearest star known, Proxima Centauri. The distinctive signature of such cool brown dwarfs is their strong methane absorption at 3.3 μm. The sensitivity must be sufficient to detect a 200 K brown dwarf at the distance of Proxima Centauri, (1.3 parsecs). The 3.3 μm channel (band 1) has been optimized to detect the methane absorption feature in brown dwarf atmospheres with T < 1000 K. The 4.7 μm channel (band 2) is optimized to provide maximum contrast with the 3.3 μm channel by centering on the nearby continuum feature in brown dwarfs.

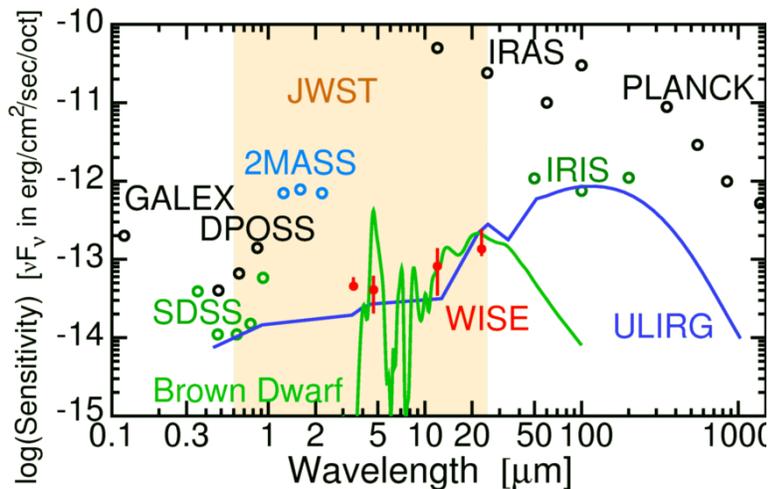

Figure 4: WISE sensitivity as it compares to various other missions and surveys.

WISE will achieve a signal to noise ratio of 5 or more in band 2 on point sources with fluxes of 0.16 mJy in regions unconfused by Galactic sources, but including the effects of extragalactic source confusion. The expected flux at 4.7 µm (a wavelength band unobscured by methane absorption) from a 200 K brown dwarf at 1.3 parsec is 0.22 mJy, enabling a 7 sigma detection and construction of a 5 sigma color limit given a 3.3 µm limit at 7 sigma. WISE shall achieve a signal to noise ratio of 5 or more in band 1 on point sources with fluxes of 0.12 mJy in regions unconfused by Galactic sources, but including the effects of extragalactic source confusion. We derive this requirement based on the goal identifying brown dwarfs cooler than 1000 K, the temperature of Gliese 229B. The 3.3 µm flux is 0.17 mJy for a 1000K blackbody with a 4.7 µm flux of 0.22 mJy, enabling a 7 sigma limit and construction of a 5 sigma color limit given a 4.7 µm detection at 7 sigma.

## 2.2. Ultraluminous Galaxies

A major goal for WISE is to identify the most luminous galaxies in the universe. The 12 µm channel (band 3) is optimized to enhance the detection of ultraluminous galaxies with redshifts up to 3, when combined with the 23 µm channel (band 4). Such galaxies are expected to require assembly until more recent epochs than redshift $z=3$. The most luminous galaxy known, the $z=0.93$ source FSC15307+3253, would have a flux of 2.6 mJy at 23 µm if it were at $z=3$. Thus, we require WISE to achieve a signal to noise ratio of 5 or more in band 3 on point sources with fluxes of 0.65 mJy in regions unconfused by Galactic sources, but including the effects of extragalactic source confusion. Such galaxies can have a 12 to 23 µm color in the Vega system as large as 3, corresponding to a flux ratio of 4 and thus a 12 µm flux of 0.65 mJy for a source with 2.6 mJy at 23 µm. We therefore require WISE to achieve a signal to noise ratio of 5 or more in band 4 on point sources with fluxes of 2.6 mJy in regions unconfused by Galactic sources, but including the effects of extragalactic source confusion.

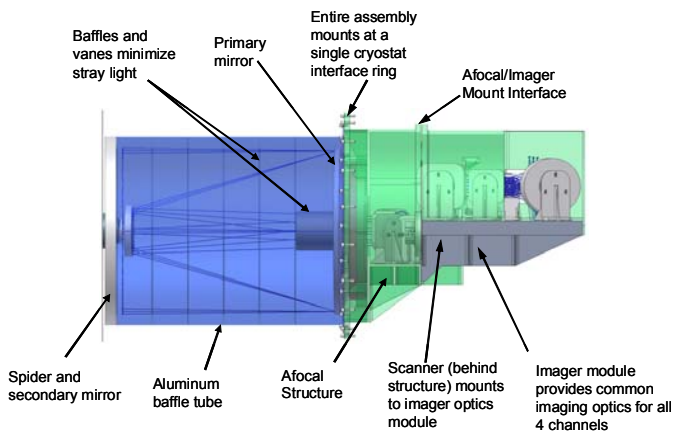

Figure 4: The WISE optical system, including the afocal telescope and MWIR and LWIR imagers. Figure from Schwalm et al. 2005.

## 2.3. Bandpass Optimization

A study was performed to optimize the four WISE bandpasses to improve sensitivity and contrast between our desired science targets (Mainzer et al. 2005). This study incorporated the latest results from the Spitzer Space Telescope to better improve our understanding of the methane feature in brown dwarfs, the zodiacal background, the spectra of ultraluminous infrared galaxies, and the spectra of other scientifically interesting targets. The results of this optimization significantly improved sensitivity margins and more tightly specified out-of-band blocking to avoid contamination due to filter leaks.

## 3. The WISE Payload

The WISE instrument will operate in a single mode, continuously scanning the sky as the sun-synchronous, 500 km altitude orbit precesses around the celestial sphere in 6 months (Larsen & Schick, 2005). The focal planes, with $1024^2$ pixel arrays and 2.75 arcsec pixels, will cover a 47 arcmin field of view. The cryogenic instrument includes a scan mechanism to offset the orbital motion and freeze the sky on the arrays. The orbit precesses in ecliptic longitude by 1°/day. WISE has a 40-cm-diameter afocal telescope, a scan mirror, and two 2-band reimaging cameras, all contained within a two-stage solid-hydrogen cryostat. Thus, the WISE telescope is launched cold. The cryostat is based on the WIRE design with relaxed temperature requirements. The WISE cryostat achieves a 7-month lifetime with a 130% margin. The scan mirror uses a flight-proven, redundant design from the SPIRIT III/MSX program and is driven in a sawtooth pattern to cancel the motion of the satellite for the 8.8 s necessary to expose the arrays. This allows 1.1 s for the flyback between exposures, for a total of 11 s between exposures. Should the mechanism fail, conventional "point and shoot" operation will still achieve the minimum mission science objectives. The end-to-end image quality, including the effects of attitude control errors and pixel size, meets the 12 μm diffraction limited requirement with margin over the full 47 arcmin FOV on the sky.

### 3.1. Focal Plane Arrays

The 3.3-μm and 4.7 μm focal plane arrays (FPAs) will use a Rockwell HAWAII 1-RG HgCdTe arrays with a 5 μm cutoff. The 12-μm and 23-μm channels will use Si:As BIB detectors on a new multiplexer designed for low read noise below 10 K. The Si:As arrays are being manufactured by DRS of Anaheim, CA. The multiplexer for the $1024^2$ Si:As arrays represent a new development; to mitigate risk, early fabrication was initiated at DRS. The new readout approximates the layout of the HAWAII-1RG as closely as possible, supporting common interface cabling, with bond pads at one end. The first multiplexer lot run has been completed and hybridized to suitable detector material (see Figure 5). Initial tests of the new hybrids have shown flight-quality performance; the parts are sufficient quality to make a second lot run unnecessary. Initial testing of the detectors indicates satisfactory read noise, dark current, and quantum efficiency. DRS will also provide the FPA mounts and electronics to the WISE payload. Table 3 summarizes the performance characteristics of the Si:As and HgCdTe FPAs. A high-background version of the WISE 1024x1024 Si:As array suitable for ground-based use is being developed by JPL and DRS (Mainzer et al. 2005)

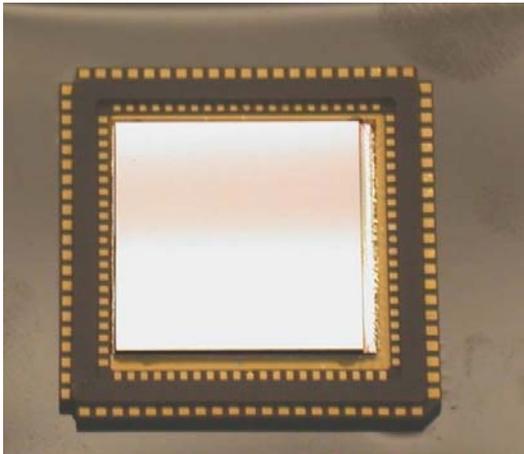

**Figure 5: The new WISE 1024x1024 Si:As hybrid. Flight quality devices were produced in the initial foundry run.**

Although the HgCdTe arrays are off-the-shelf devices, a critical modification was necessary to make them compatible with the space environment. The HgCdTe layer, which is ~5 μm thick, is grown on top of a thick (~800 μm) CdTe substrate. Other NASA programs (including the NIRCAM and NIRSPEC instruments for JWST and the Wide Field Camera for Hubble) found that the CdTe layer suffers from photoluminescence when the array is bombarded with protons (Johnson et al. 2004, McKelvey et al. 2004, Hill et al. 2005). The photoluminescence manifested both as a short-term (immediate) effect and a longer term persistence. Dr. Craig McCreight served on our WISE detector Preliminary Design Review board, and he alerted us to the problem, which they had identified in the course of testing HgCdTe arrays for JWST. When we examined dark data from Dr. McCreight's group, we found that each proton hit caused "clouds" of several thousand pixels to be illuminated – clearly unacceptable given the predicted WISE fluence of 1 hit/cm$^2$/sec. This fluence is expected to produce ~40-50 cosmic ray hits per frame for WISE; if each hit were to produce several thousand pixels each several sigma above the nominal dark level, the project would be unable to meet photometric accuracy requirements. This short term photoluminescence is caused by the production of blue photons in the CdTe layer, most likely due to the high cross-section of the Te. Based on this analysis, the decision was made by the WISE project to remove the CdTe layer. All other space-based projects using HgCdTe grown on CdTe substrates are removing the substrate layer as well. Removing the layer is a relatively well-understood process that is accomplished with the aid of an etch stop layer grown between the HgCdTe and the CdTe. Rockwell predicts that removing the layer will result in minimal yield losses and little change in performance. This decision puts us in line with the best practices of other NASA infrared missions. We are carefully tracking the HgCdTe results from our sister missions.

| Parameter | Si:As Requirement | Si:As Predicted Performance | HgCdTe Requirement | HgCdTe Predicted Performance |
|---|---|---|---|---|
| FPA Format | 1024 x 1024 pixels | | 1024 x 1024 pixels | |
| Operating Wavelengths | 7.5 - 16.5 µm and 20 – 28 µm | | 2.8 – 3.8 µm and 4.1 – 5.2 µm | |
| Pixel Pitch | 18 µm | | 18 µm | |
| Pixel Operability | >90% | | > 90% | |
| Bad Pixel Clustering | No clusters of > 16 individual bad pixels | | No clusters of > 16 individual bad pixels | |
| Read Noise for a single 7 sample operation | < 40 e rms | 42 e rms (measured) | ≤ 7 e rms | 15 e rms (measured) |
| Dark Current @ operating temp | < 100 e/sec | < 5 e/sec (measured) | < 1 e/sec | |
| Mean Quantum Efficiency over 7.5 - 16.5 µm and 20-28 µm with AR coating | > 0.5 | > 0.7 | > 0.5 | > 0.7 |
| Well Capacity | > $10^5$ e | >$10^5$ e | >$10^5$ e | >$10^5$ e |
| Electrical Crosstalk | ≤ 1% | ≤ 1% | ≤ 1% | |
| Optical Crosstalk | ≤ 5% | ≤ 5% | ≤ 5% | |
| Array Readout Time | < 1.1 sec | | < 1.1 sec | |
| Reset Time (includes read) | < 1.1 sec | Row reset – 2 msec | < 1.1 sec | |
| FPA Temperature | 7.8 ± 0.5 K | < 9 K | 30 K | |
| Reference Pixels | ≥ 4 pixel-wide band around active pixel area | | ≥ 4 pixel-wide band around active pixel area | |
| Heat Dissipation | < 4 mW each FPA including heat conduction through cables | < 3 mW each FPA | < 8 mW each FPA including heat conduction through cables | < 3 mW each FPA including heat conduction through cables |

**Table 3: Si:As and HgCdTe FPA key parameters.**

### 3.2. Optical Design

The WISE instrument consists of a 40-cm baffled telescope which uses three dichroic beamsplitters to illuminate four FPAs in bands centered around 3.3, 4.7, 12, and 23 µm (Schwalm et al. 2005). A cryogenic scan mirror is used to stabilize the image against the orbital rotation (see Figure 6). An engineering model of the scan mirror mechanism has recently been completed and is undergoing test at SSG. The WISE optical system consists of the afocal five mirror telescope assembly with 40 cm primary mirror, the refractive MWIR imager and reflective LWIR imager. The instantaneous field of view is 47 x 47 arcmin, with 2.75 arcsec pixels. Figure 4 provides a block diagram of the WISE optical system. The afocal design allows for more compact packaging and minimal distortion across the field of view. Both the MWIR and LWIR imagers will have all reflective elements to minimize reliance on exotic materials. The four focal planes share the same field of view, with three beamsplitters separating the light first into the MWIR and LWIR channels, then two more splitting the light into the four individual arrays. This optical layout allows the individual modules to be assembled and tested independently, allowing good clearance for the scan mirror mechanism. All four focal planes are located near the back of the cryostat, allowing for convenient mechanical and electrical access. Figure 7 depicts the optical trains for MWIR and LWIR imagers. A system model including PSF, wavefront error, spacecraft jitter, etc. was created to form a performance metric. The fully-assembled WISE instrument will be characterized on the ground using special purpose calibration facilities. The final calibration will be derived from on-orbit observation of selected celestial sources during IOC.

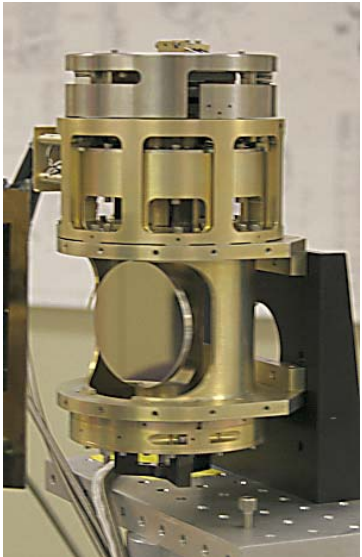

**Figure 6: The engineering unit WISE scan mirror mechanism.**

### 3.3. Cryogenic Support Subsystem

The cryogenic support system consists of a solid hydrogen cryostat, deployable cover, dual-stage aperture shade and thermally isolating support structure to the spacecraft (Naes et al. 2005). The WISE cryostat is a dual-stage, $SH_2/SH_2$ cryostat designed for a 7-month mission, with a 130% lifetime margin. The primary tank will operate at ~6.5K and will cool the 12 and 23μm focal planes to 7.8K±0.5K with a total heat load of 12mW. To achieve this low operating temperature, the <12K secondary tank acts as a guard to the primary tank, and intercepts essentially all parasitic heat loads from the ambient structure. The secondary also provides instrument cooling of the telescope and scan mirror to <17K and acts as the heat sink for the 32K, band 1 and 2 HgCdTe focal planes. Two vapor-cooled shields, mounted intermediately along the support structure, use the hydrogen effluent vapor to absorb a significant portion of the incoming parasitic heat. To close out the vacuum space, a deployable dome-shaped aperture door assembly is provided. In a mission critical event during IOC the aperture door is spring-ejected following the release of three NSI-actuated separation nuts. The design has fully independent arm and fire mechanisms. The inner shield of the aperture door can be cooled with liquid helium using an open-loop heat exchanger circuit. Cooling this inner shield will allow for full instrument functional checkouts while the instrument is on the ground. An aperture shade is mounted at the telescope entrance to insulate the open cryostat system from direct Earth and Sun heat loads after the cover ejection.

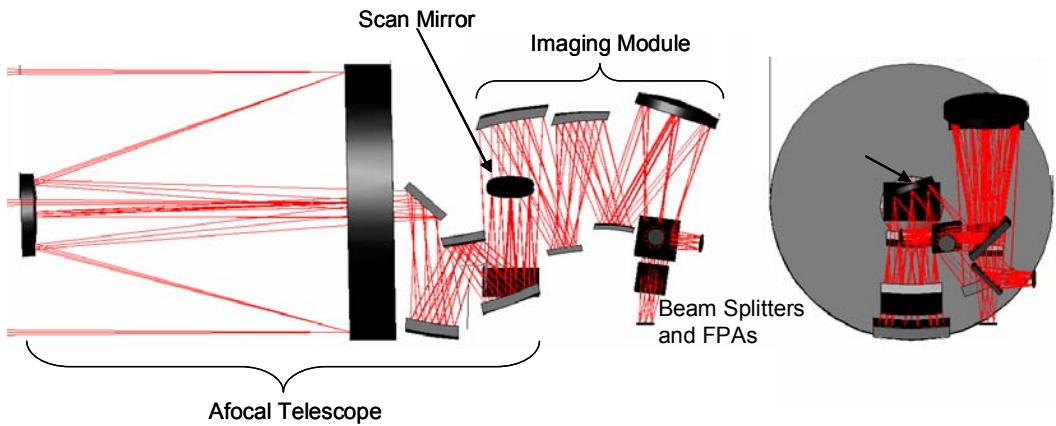

**Figure 7: The MWIR and LWIR imagers. Figure from Schwalm et al. 2005.**

## 4. The WISE Spacecraft

The WISE Observatory will provide five fundamental operational modes: Science Survey Mode, in which the instrument takes images continuously every 11 sec while the spacecraft maintains a continuous pitch rate that matches the orbit pitch, Uncompressed Science Survey Mode, in which compression is turned off, as well as downlink, safe, and emergency modes.

WISE will adapt the single string Ball Aerospace and Technologies Corp. (BATC) RS300 spacecraft bus by increasing its size to match the Delta, and upgrading the telecommunications system. The fixed high-gain antenna (HGA) is based on the Deep Impact design. Attitude control to 50" is all that is required; precision attitude knowledge will come from ground science data analysis. The spacecraft attitude jitter during an 8.8 sec exposure is 0.54" (1-σ, per axis), well within the 2.75" pixel size. The solar array for WISE is fixed and provides 467 W end-of-life power. Electrical power is generated by 3.0 m² of 28.0% efficiency triple-junction solar cells which are fixed to the spacecraft body.

The command and data handling (C&DH) system resides in the Spacecraft Control Avionics (SCA). The SCA provides a platform for the flight software and power distribution interfaces for the Observatory and the attitude determination and control subsystem (ADCS) components. The C&DH portion of the SCA contains the single-board computer, nonvolatile memory, command and telemetry interface, 1553B interface, and the instrument interface. It contains a RAD750 single board computer. The WISE instrument will transfer the data to the SCA, where hardware will perform lossless Rice 2.1:1 compression on the science data.

The telecommunications subsystem is designed to be used to support communications with the Tracking Data Relay Satellite System (TDRSS). The WISE bus will be configured to support a 120-Mbps Ku-band single access return link for all science data downloading. The bus will also be configured to support forward and return S-band link to provide command and housekeeping telemetry downloads. Data storage will be provided by a Redundant Array Of Independent Disk Drives (RAID), with 85.9 GB of storage for the mission.

The ADCS architecture is a three-axis stabilized system. A precision star tracker and inertial reference unit with three reaction wheels provide precise attitude determination and control. Magnetometers and torque rods with redundant windings are used for wheel momentum management. Coarse Sun sensors determine the Sun vector in any spacecraft orientation. Three reaction wheels allow the spacecraft accurate pointing and steady slewing. The momentum will be dumped by the torque rods above 60° latitude during science data downlinks.

## 5. Mission Operations Concept

The goal of the WISE data concept is to provide a lasting legacy to the astronomical community in the form of a permanent whole sky source catalogue and image atlas in the mid-infrared (3.3 to 23μm). These products will be made accessible to the community in collaboration with the NASA Infrared Science Archive (IRSA) at the Infrared Processing and Analysis Center (IPAC), thus ensuring long-term availability of these products beyond the end WISE missions operations and data processing phase. As a baseline data will be made available to the community via the internet. The data will be maintained in a way that distribution of the complete WISE source catalogue via Digital Versatile Disk (DVD) to frequent users is possible. All image data will be made available in accordance with the Flexible Image Transport (FITS) astronomical data standard, tables will be in the widely used IPAC table format. In order to ensure survivability in case of a major catastrophe causing the loss of the data at the IPAC facility a complete copy of the WISE data set and software source code will be deposited at a secure off-site location.

A fundamental objective of the WISE project is to make data products available to the astronomical community as soon as is technically feasible and scientifically sensible. To facilitate this, WISE data products will be released in two stages. These will occur 6 and 17 months after the end of the nominal on-orbit lifetime of 7 months. The first data release will allow immediate use of WISE data by the community, and will consist of a preliminary image atlas and source catalog containing sources that have SNR of 20 or higher in unconfused regions of sky. The final data release will include sources to about SNR = 5 and will be accompanied by more extensive quality analysis and product validation.

### 5.1. WISE Mission Operations System

The WISE Mission Operations System is divided into three subsystems, summarized in Figure 9. The PI team at UCLA provides the scientific survey planning for the mission. On a weekly basis it generates the spacecraft pointings and scan parameters to execute an optimized survey plan allowing for SAA compensation, lunar avoidance and if necessary recovery of gaps resulting from inadvertent data losses, e.g. after a safe mode recovery. This leads to a nominal survey plan providing a coverage of better than 99.25% of the sky with more than 8 independent exposures during the 6 months prime mission.

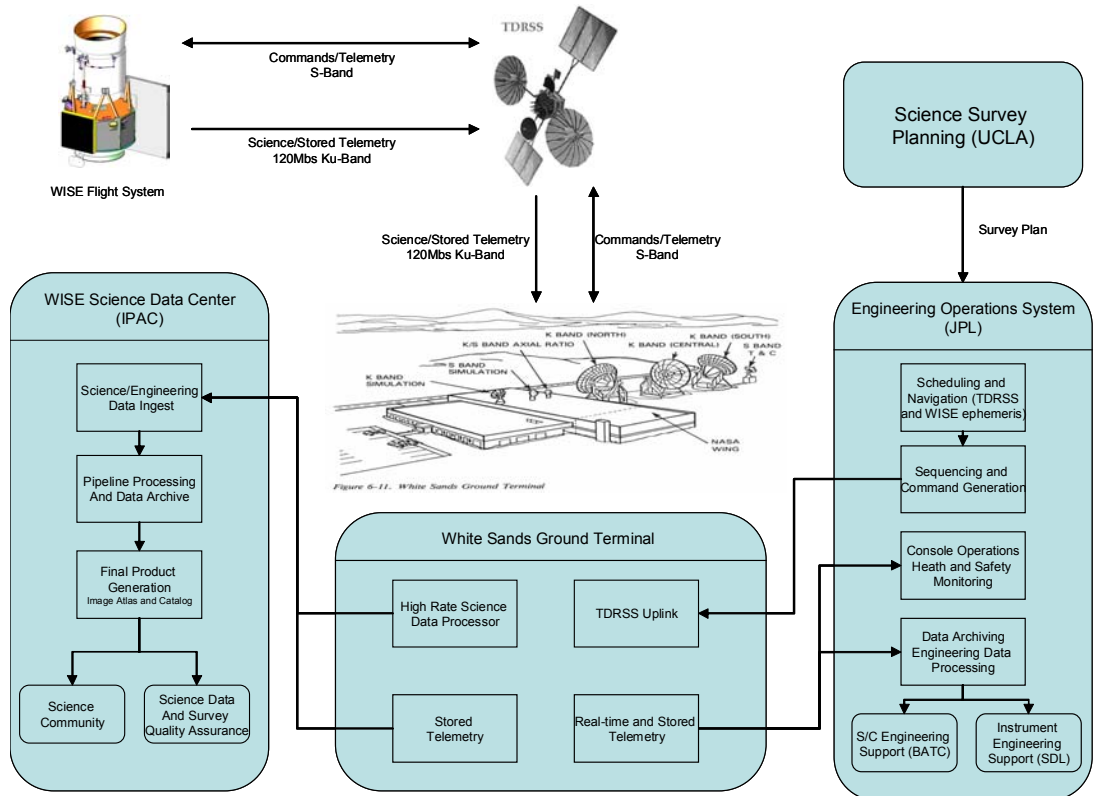

**Figure 9: The WISE Mission Operations System, include flight and ground segments.**

Engineering operations for the mission are provided by the JPL Earth Science Mission Center, which is currently operating the Jason and Topex spacecrafts. The Engineering Operations (EOS) team of JPL engineers and operators is supported by experts from both the spacecraft (BATC) and the science instrument (SDL) providers forming an integrated spacecraft team during on-orbit operations. The EOS is responsible for the health and safety of all WISE mission operations. It will monitor the flight system and perform all necessary spacecraft and payload maintenance operations. It provides the detailed scheduling and navigation function and generates the weekly sequence loads to WISE via the TDRSS relay satellite system.

At the beginning of each TDRSS communication pass the EOS real-time operator will establish the heath and safety status of the WISE flight-system by checking the status pages for any red or yellow alarm violations or unexpected flight-system configuration. In addition to this health and safety check, the housekeeping telemetry accumulated during each period of autonomy will undergo analysis by the EOS spacecraft team to identify any possible transient anomalies or unexpected flight-system behavior. The EOS also provides the first step in the processing of the high rate science data when it is received via TDRSS. It will frame sync, RF decode and extract telemetry packets out of the high rate data stream, followed by packet level data accountability, which will be automatically evaluated based on pre-set thresholds for required data re-transmissions. Subsequently it will transfer the packet data via data-lines to the WISE Science Data Center at IPAC for further processing.

## 5.2. Data Reduction and Archiving

WISE will provide a lasting legacy to the astronomical community in the form of a photometrically and astrometrically calibrated digital Image Atlas covering the entire sky in the four survey bands, and a Catalog containing accurate positions and brightnesses for all sources extracted from the image data. The WISE Science Data Center (WSDC) located at the Infrared Processing and Analysis Center (IPAC) is responsible for converting the raw imaging data from the spacecraft into final Image Atlas and Source Catalog, and to archive and distribute those products.

WISE data processing is organized into four basic functions, each handled by a separate subsystem at the WSDC: Ingest, Data Reduction Pipelines, Final Product Generation, and Archiving.

a. **The Ingest subsystem** autonomously receives raw science and engineering data that are transmitted to the WSDC from the White Sands facility and JPL/EOS, respectively, following each downlink pass. The science and engineering data are decompressed and merged to form sets of raw images in Flexible Image Transport System (FITS) format corresponding to each WISE exposure on the sky.

b. The **Data Reduction Pipelines** subsystem is a highly automated, high-throughput dual-stage software system that is based on the architecture of the processing system used to reduce data from the Two Micron All-Sky Survey (2MASS; Cutri et al. 2003). Data from individual orbits are first reduced as a unit in the "single-orbit pipeline." Single-orbit processing steps include derivation and application of instrumental calibration corrections (flat-field, linearity, sky illumination, bad pixel masks), source detection and characterization (photometry and astrometry) on individual image frames, merging source detections from the four bands, artifact detection flagging, and identification of known solar system objects in the extracted source lists. Photometric calibration of single-orbit data is derived from measurements of a network of standard stars located near the ecliptic poles, and astrometric transformations are determined using the measurements of 2MASS Point Source Catalog stars in each image. Once data from a set of adjacent orbits have been processed, the "multi-orbit" pipeline is run. This software system determines the relative position offsets between adjacent orbit images, shifts and registers the images onto a common spatial grid, then smooths, resamples and coadds the images to generate the full-depth Atlas Images and depth-of-coverage maps. Source detection and characterization is then run on the combined images, the detections from the four bands are merged, detections of image artifacts are identified and flagged, and the final source extractions are output to a source *Working Database*.

c. The **Final Product Generation** subsystem draws the subset of entries from the extracted source *Working Database* and archive of Atlas Images that constitute the WISE Source Catalog and Image Atlas. This is an iterative process that determines the set of selection criteria that generate products that meet the reliability, completeness, photometric and astrometric accuracy of the mission. Final Product Generation also entails validating the resulting products before their release.

d. The **Archive** subsystem at the WSDC stores raw and processed survey data, and provides a "living" archive of WISE data products that enables distribution to the WISE project, astronomical community and the public. Raw telemetry is stored on local magnetic disk for the duration of the mission, and written to magnetic tape for long-term archiving. The processed image and extracted source data from both the single- and multi-orbit pipelines are integrated into the DBMS systems of the Infrared Science Archive (IRSA) located at IPAC. Access to the processed data products by the WISE team during mission operations, and by the public to the WISE Catalog and Image Atlas is via the web-based IRSA tools. These include image preview, retrieval and inventory services, powerful catalog and database query engines, as well as services that provide interoperability with other IRSA catalog and image holdings. Release of the WISE data products through IRSA insures accessibility to the Virtual Observatory (VO) because all IRSA services are compliant with current VO protocols and standards, and are registered in the US National Virtual Observatory.

## 6. Project Status

The Wide-field Infrared Survey Explorer (WISE) project is an element of the NASA Explorer Program managed by Goddard Space Flight Center. The WISE Project has completed an extended Phase A and entered Phase B in October, 2004. Preliminary Design Reviews for all major subsystems, including the payload, spacecraft bus, and mission operations have been successfully completed. The mission Preliminary Design Review is scheduled for late July, 2005. The WISE Project will stand before a NASA Confirmation Review board in October of 2005, requesting approval to move into Phase C/D.

The project is led by its University of California at Los Angeles Principal Investigator, Dr Ned Wright. Jet Propulsion Laboratory will manage the project and provide System engineering leadership; Ball Aerospace and Technology Corporation (BATC) will provide the spacecraft, lead flight system test and support launch operations; Utah State University's Space Dynamics Laboratory (SDL) will provide the Payload; operations will be led by JPL with science data processing and archiving by Caltech's Infrared Processing and Analysis Center (IPAC) using NASA's TDRSS facility for commanding and data retrieval; education and public outreach will be provided by the University of California at Berkeley.

Key project milestones which are controlled by this project plan are listed in Table 4. Dates are contingent on timely confirmation decisions between Phase B and C/D, E.

| Phase B Start | October, 2004 |
|---|---|
| SRR | December, 2004 |
| PDR | July, 2005 |
| CDR | December, 2006 |
| Payload delivery | November, 2008 |
| Observatory delivery to WTR | March, 2009 |
| Launch | June, 2009 |
| IOC complete | L+ 1 M |
| End of Survey (EOS) | L + 7 M |

The project is currently completing the preliminary design and final cost estimates in preparation for its confirmation review. Testing of the flight detectors was performed this year with satisfactory results. An engineering model scan mirror has been completed and is undergoing test. Further information on WISE can be found at http://wise.ssl.berkeley.edu/.

## 7. Acknowledgements

**Table 4: Key project milestones.**